\documentclass[aps,prb,superscriptaddress,amsmath,amssymb,floatfix,twocolumn]{revtex4-1}
\usepackage{graphicx}
\usepackage{subfigure}
\usepackage{color}
\usepackage{epstopdf}
\usepackage{mathrsfs}
\newcommand{\mbS}{\mathbf{S}}
\newcommand{\bdd}{\boldsymbol{\delta}}

\begin{document}

\title{Two-magnon Raman scattering in antiferromagnetic phases of frustrated spin models on the honeycomb lattice}

\author{Junru Pan}
\affiliation{Department of Physics and Beijing Key Laboratory of Opto-electronic Functional Materials and Micro-nano Devices,
Renmin University of China, Beijing 100872, China}

\author{Feng Jin}
\affiliation{Beijing National Laboratory for Condensed Matter Physics, Institute of Physics, Chinese Academy of Sciences, Beijing 100190, China}

\author{Jianting Ji}
\affiliation{Beijing National Laboratory for Condensed Matter Physics, Institute of Physics, Chinese Academy of Sciences, Beijing 100190, China}

\author{Qingming Zhang}
\affiliation{School of Physical Science and Technology, Lanzhou University, Lanzhou 730000, China}
\affiliation{Beijing National Laboratory for Condensed Matter Physics, Institute of Physics, Chinese Academy of Sciences, Beijing 100190, China}

\author{Rong Yu}
\email{rong.yu@ruc.edu.cn}
\affiliation{Department of Physics and Beijing Key Laboratory of Opto-electronic Functional Materials and Micro-nano Devices,
Renmin University of China, Beijing 100872, China}

%\date{\today}

\begin{abstract}
We calculate the two-magnon Raman scattering spectra in antiferromagnetic phases of several frustrated spin models defined on the honeycomb lattice. These include the N\'{e}el antiferromagnetic phase of a $J_1$-$J_2$-$J_3$ model and the stripe phase of the Heisenberg-Kitaev model. We show that both the magnetic frustration and the anisotropy of interactions may significantly affect the Raman spectra. We further discuss the implications of our results to the magnetic excitations of the iron-based compound BaFe$_2$Se$_2$O and show how the magnetic interactions can be extracted from fit to the Raman spectrum.
\end{abstract}

\maketitle

%\onecolumngrid

\section{Introduction}

Frustrated magnets constitute one of the most active subjects in condensed matter physics~\cite{Diep_Book,Lacroix_Book}. These systems can hold exotic states of matter and may exhibit novel quantum behaviors, such as spin skyrmion lattice~\cite{Villalba_PRB2019} and quantum spin liquids (QSLs)~\cite{Savary_RoPP2017,Zhou_RMP2017,Chamorro_CR2021}. In frustrated magnets, systems with geometric frustration, such as those defined on the triangular or Kagome lattices, have been extensively studied, and a number of candidate QSL systems have been proposed~\cite{Fu_Science2015,Shen_Nat2016}.

Recently, the honeycomb lattices antiferromagnets attract great interest. Though the honeycomb lattice is bipartite and the Heisenberg antiferromagnetic (AFM) model with the nearest neighbor interaction defined on it has a N\'{e}el order, exchange couplings between further neighboring spins and/or anisotropic interactions may still introduce strong frustration and cause highly nontrivial effects. For example, a valence bond solid state is proposed when next-nearest neighbor interactions or ring-exchange exchange couplings are considered~\cite{Albuquerque_PRB2011,Ganesh_PRL2013,Pujari_PRB2015}. A QSL has been found in the exactly solvable Kitaev model~\cite{Kitaev_AP2006}. Even in the classical spin systems, exotic magnetic structures such as zigzag, stripe, or spiral orders, appear as a consequence of the frustration effect~\cite{Rastelli_Physica1979,Fouet_EPJB2001}.

Experimentally, many honeycomb lattice antiferromagnetic compounds have been synthesized. These materials exhibit interesting features. On the one hand, most of them are magnetically ordered at low temperatures. For example, the quasi-two-dimensional BaNi$_2$V$_2$O$_8$ shows a N\'{e}el order~\cite{Rogado_PRB2002}, while the isostructural compound BaCo$_2$As$_2$O$_8$ exhibits a spiral order~\cite{Regnault_Physica1977}. The zigzag order was found in the Kitaev model candidate compound $\alpha$-RuCl$_3$,\cite{Sears_PRB2015} and the stripe AFM order is stabilized in Ba$_2$NiTeO$_6$.\cite{Asai_PRB2017} The very recently reported rare-earth Kitaev material YbOCl is also found to be magnetically ordered~\cite{Luo_SP2020,Ji_CPL2021}. On the other hand, the honeycomb lattice is usually distorted in these materials. For instance, both $\alpha$-RuCl$_3$ and Mn$_2$V$_2$O$_7$ show a monoclinic distortion which reduces the symmetry group to $C2/m$.\cite{Kim_PRB2016,Sun_MRE2017} The lattice distortion could be crucial for the stabilization of the magnetic order, as it may provide a way to release the spin frustration.

As one major theme of studying the honeycomb lattice antiferromagnets is to seek the QSL states~\cite{Takagi_NRP2019}, it is then very important to understand the effects of lattice distortion and anisotropy of exchange interactions to the magnetism of the system. In this way, one finds how the system can be tuned to the possible QSL regime of the magnetic phase diagram. However, to know how faraway the experimental system is to the regime of interest in the phase diagram, one needs to determine the exchange couplings of the given material. Unfortunately, this is usually a hard task for physicists. Though the order of magnitude of the exchange interactions can be easily estimated from the susceptibility or specific heat data, to obtain more precise values of competing (and frustrated) interactions one usually needs a decent fit to inelastic neutron scattering spectrum at low temperatures. Sometimes this is unavailable, and the reason is two-fold: On the theory side, the understanding on the low-temperature spin excitations of the system is often limited, and on the experimental side, main issues could be the lack of qualified samples and appropriate experimental conditions.

Raman scattering is a unique and powerful tool for probing the spin excitations~\cite{Lemmens_PR2003,Devereaux_RMP2007}. Compared to neutron scattering, though the momentum information is lacking, the Raman scattering may offer a higher energy resolution. For a magnetically ordered system described by a local spin Hamiltonian, the Raman scattering probes two-magnon correlations in which short-wavelength excitations dominates. A standard magnetic Raman scattering theory in a spin system is based
on the Fleury-Loudon (FL) coupling between the light and the spin system~\cite{Fleury_PRL1967,Fleury_PR1968}, which was derived from the Hubbard model in the large-$U$ limit~\cite{Shastry_PRL1990}. Giving these, it is very intriguing to extract the exchange couplings of the system by fitting theoretical results to the measured Raman spectrum.

In this paper we study the effects of anisotropy of interactions and magnetic frustration on the Raman scattering spectra in several antiferromagnetically ordered states on the honeycomb lattice. We show that both the interaction anisotropy and magnetic frustration can significantly tune the bandwidth of the two-magnon Raman scattering spectrum. We then discuss the implications of our results to the magnetic excitations of the iron-based compound BaFe$_2$Se$_2$O.\cite{Han_PRB2012,Popovi_PRB2014,Jin_PRB2019} This gives an example on how the magnetic interactions can be extracted from fit to the Raman spectrum. The rest part of the paper is organized as follows: In Sec.~\ref{Sec:Model}, we introduce the anisotropic AFM honeycomb lattice models and the formulation for calculating the two-magnon Raman spectrum in the AFM ordered states. In Sec.~\ref{Sec:ResultNeel}, we present the results of calculated Raman spectrum in the anisotropic $J_1$-$J_2$-$J_3$ model and show that both the spatial anisotropy of interaction and magnetic frustration lead to narrowing of the bandwidth of the spectrum, while originating from different mechanisms. In Sec.~\ref{Sec:ResultKitaev}, we further show the Kitaev interaction can also strongly modify the two-magnon Raman spectrum and suppresses the overall spectral weight. In Sec.~\ref{Sec:Fe}, we calculate the two-magnon spectrum for the iron-based compound BaFe$_2$Se$_2$O and compare our results to the experimental one. We show that the exchange interaction of the material can be obtained from the fitted peak energies of the spectrum. The determined exchange couplings differ largely from those in the first-principles calculation, and imply completely low-energy physics for this compound. Finally, concluding remarks are presented in Sec.~\ref{Sec:Conclusion}.

\section{Model and method}\label{Sec:Model}
\begin{figure}[t!]
\centering\includegraphics[%scale=0.28
width=80mm %, trim=5 130 5 20,clip %left, down, right, up
]{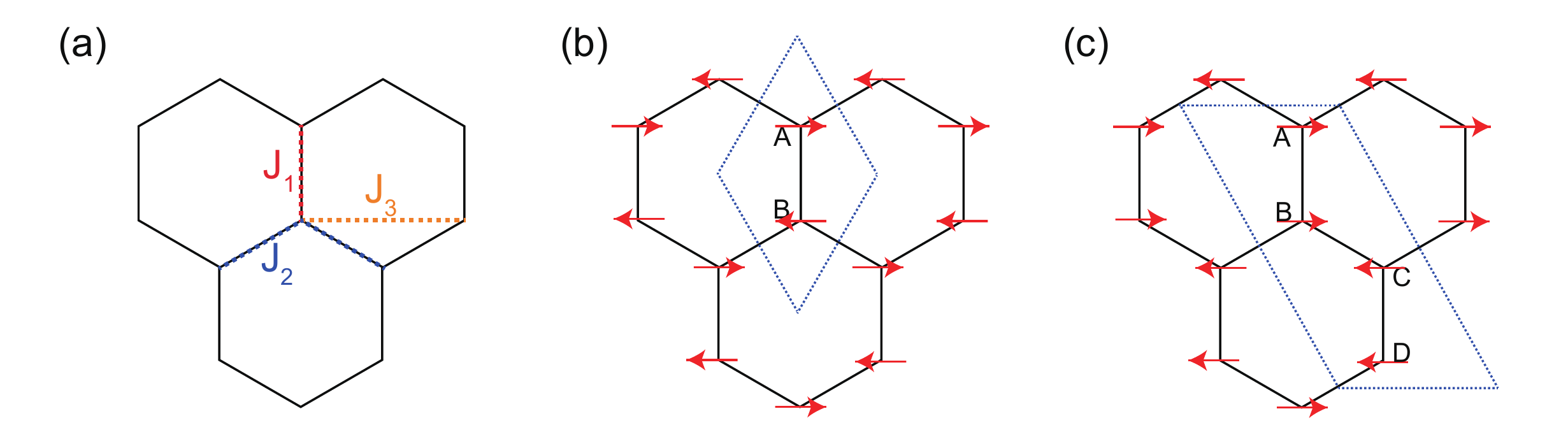}
\caption{(a): Sketch of the honeycomb lattice and corresponding exchange couplings $J_1$, $J_2$, and $J_3$ described in Eq.~\eqref{Eq:M1}. (b): The N\'{e}el AFM spin configuration. (c): The stripe AFM spin configuration.}
\label{fig:1}
\end{figure}

In this paper we consider two variations of the AFM Heisenberg model defined on the honeycomb lattice. The first one is a spatially anisotropic $J_1$-$J_2$-$J_3$ Heisenberg model. The Hamiltonian reads as
\begin{align}\label{Eq:M1}
H=&J_1\sum_{i,\bdd_1=\vec{a}_1}\mbS_{A,i}\cdot\mbS_{B,i+\bdd_1} +J_2\sum_{i,\bdd_2=\{\vec{a}_2,\vec{a}_3\}}\mbS_{A,i}\cdot\mbS_{B,i+\bdd_2} \nonumber\\ +&J_3\sum_{i,\bdd_3=\vec{b}_1}\mbS_{A(B),i}\cdot\mbS_{A(B),i+\bdd_3},
\end{align}
where $\mbS_{A/B,i}$ refers to the spin operator at site $i$ of the $A/B$ sublattice, $\vec{a}_j$ $(j=1,2,3)$ are the lattice vector along the three nearest-neighbor directions, and $\vec{b}_1$ is the lattice vector along the next nearest-neighbor direction that perpendicular to $\vec{a}_1$. $J_1$ and $J_2$ are the corresponding nearest-neighbor AFM exchange couplings, and $J_3$ refers to the next-nearest-neighbor AFM exchange coupling (Fig.~\ref{fig:1}(a)). For general exchange coupling values, the model is spatially anisotropic which reflects the lattice distortion in the honeycomb system, and the Hamiltonian has the point group symmetry of $D_{2h}$. In addition, the next-nearest-neighbor AFM coupling $J_3$ introduces spin frustration. Note that when $J_3=0$ and $J_1=J_2$, the system recovers the isotropic and non-frustrated honeycomb lattice AFM Heisenberg model, with the point group symmetry $D_{3d}$. This non-frustrated model has a N\'{e}el AFM ground state as shown in Fig.~\ref{fig:1}(b). This ground state is also stabilized when the anisotropy and frustration of the interactions are moderate.

The other model is the Heisenberg-Kitaev model with the nearest-neighbor interactions. The Hamiltonian reads as
\begin{align}\label{Eq:M2}
H=&\sum_{i,\bdd=\{\vec{a}_1,\vec{a}_2,\vec{a}_3\}}\left\{J\mbS_{A,i}\cdot\mbS_{B,i+\bdd_1} %\nonumber\\
-K%\sum_{i,\bdd=\{\vec{a}_1,\vec{a}_2,\vec{a}_3\}}
S^{\gamma_{\bdd}}_{A,i}\cdot S^{\gamma_{\bdd}}_{B,i+\bdd}\right\},
\end{align}
where $S^{\gamma_{\bdd}}=S^x,S^y,S^z$ refer to the three spin components corresponding to the bond direction $\bdd$, and $K$ is the ferromagnetic (FM) Kitaev coupling. With this interaction, the associated point group symmetry is still $D_{3d}$. It has been shown that the Kitaev coupling will drive the system from the N\'{e}el AFM state to a stripe ordered state (see Fig.~\ref{fig:1}(c)).\cite{Price_PRB2013}

We consider the models with a general spin $S$, and focus on the large-$S$ limit. This applies to the iron-based compound BaFe$_2$Se$_2$O,\cite{Han_PRB2012,Popovi_PRB2014,Jin_PRB2019} which has a spin $S=2$, and is also interesting for the Heisenberg-Kitaev model~\cite{Stavropoulos_PRL2019,Consoli_PRB2020}. In the following we calculate the two-magnon Raman scattering in the N\'{e}el and stripe AFM states of models in Eqs.~\eqref{Eq:M1} and \eqref{Eq:M2} within a linear spin-wave approach according to the FL theory~\cite{Fleury_PRL1967,Fleury_PR1968}. We first introduce the Holstein-Primakoff (H-P) transformation for the spins:
\begin{align}\label{Eq:HP}
 S^z_{A,i}=&S-a^\dag_i a_i \nonumber \\
 S^+_{A,i}=&\sqrt{2S-a^\dag_i a_i} a_i \nonumber\\
 S^z_{B,j}=&b^\dag_j b_j-S \nonumber\\
 S^+_{B,j}=&b^\dag_j\sqrt{2S-b^\dag_j b_j},
\end{align}
where $a_i$ and $b_j$ refer to H-P bosons on $A$ and $B$ sublattices, respectively.
We then rewrite the Hamiltonians in terms of $a$ and $b$ bosons:
\begin{align}
 H=E^\prime_0+\sum_{k} \left[P_k(a^\dag_k a_k +b^\dag_k b_k) +Q_k(a_k b_{-k} + a^\dag_k b^\dag_{-k})\right],
\end{align}
and perform the Bogoliubov transformation to diagonalize the Hamiltonian to
\begin{align}
 H=E_0+\sum_k \omega_k (\alpha^\dag_k \alpha_k + \beta^\dag_k \beta_k).
\end{align}
In above equations, $E_0$ and $E^\prime_0$ are constants, and $\alpha_k$ and $\beta_k$ are the bosonic Bogoliubov quasiparticles with wave vector $k$. $\omega_k$ is the dispersion which depends on the coefficients $P_k$ and $Q_k$. For the N\'{e}el AFM state of the $J_1$-$J_2$-$J_3$ model in Eq.~\eqref{Eq:M1}, we find
\begin{align}\label{Eq:PQ}
 P_k =& S(J_1 + 2J_2 +2J_3(\gamma_{k3}-1) \nonumber\\
 Q_k =& S(J_1\gamma_{k1} + 2J_2\gamma_{k2}),
\end{align}
where $\gamma_{ki}=\sum_{\bdd_i} e^{\mathbf{k}\cdot\bdd_i}$.
For the Heisenberg-Kitaev model, because of the spin-orbit coupling, we need to first rotate the spins such that the $S^z$ is along the quantization axis of the corresponding magnetic order. We then perform the H-P transformation, similar to Eq.~\eqref{Eq:HP}, in the rotated spin representation.

In the FL theory, the Raman operator
\begin{align}
 \hat{O} \sim \sum_{ij} (\hat{e}_{i}\cdot\hat{d}_{ij}) (\hat{e}_{o}\cdot\hat{d}_{ij}) H_{ij},
\end{align}
where $H_{ij}$ is the local Hamiltonian between sites $i$ and $j$, $\hat{d}_{ij}$ is the vector connecting sites $i$ and $j$, and $\hat{e}_{i/o}$ is the polarization vector of the incident (outgoing) light. Now we consider a realistic setup in experiments, where the incident and outgoing lights are both along the $\hat{z}$ direction, and the polarization is along the $\hat{y}$ direction, \emph{e.g.}, $\hat{e}_{i}=\hat{e}_{o}=(0,1)$. One can easily show that the only nonzero contribution to the Raman operator comes from the $A_g$ ($A_{1g}\bigoplus E_g$) channel when the system has the $D_{2h}$ ($D_{3d}$) point group symmetry. In terms of the Bogoliubov quasiparticles, the Raman operator is expressed as
\begin{align}\label{Eq:O}
 \hat{O} \sim \sum_{k} M_k(\alpha_k\beta_{-k} + \alpha^\dag_k\beta^\dag_{-k}),
\end{align}
where $M_k$ can be obtained from $P_k$ and $Q_k$ via the diagonalization of the spin Hamiltonian, and we have neglected the constant term that has no contribution to the inelastic scattering in the above expression.

The zero-temperature Raman cross section is
\begin{equation}
R(\omega)=-\dfrac{1}{\pi}Im[I(\omega)]\label{RW}
\end{equation}
where
\begin{align}
I(\omega)&=-i\int{dte^{i\omega t}\langle \mathscr{T}_t\hat{O}^+(t)\hat{O}(0)\rangle_0}
\end{align}
is the correlation function of the Raman operator, where $\langle ...\rangle_0$ refers to the quantum mechanical average over the ground state. Using Eq.~\eqref{Eq:O},
we get the two-magnon contribution to the Raman scattering
\begin{align}
I(\omega)&\sim\sum_{k,k'}M_k\Pi_{kk'}(\omega)M_{k'}\label{IW},
\end{align}
where
\begin{align}
\Pi_{kk'}(\omega)=-i\int{dte^{i\omega t}\langle \alpha_k(t)\beta_{-k}(t)\alpha_{k'}^+(0)\beta_{-k'}^+(0)\rangle_0}
\end{align}
is the two-magnon propagator. Make use of Wick's theorem, within the spin-wave approximation, we obtain
\begin{align}
\Pi_{kk'}(\omega)
&=\delta_{kk'}\dfrac{1}{\omega+i0^+-2\omega_k},
\end{align}
which leads to
\begin{align}\label{Eq:Rexp}
R(\omega)&=-\dfrac{1}{\pi}Im[I(\omega)]\notag\\
&=\sum_k M_k^2\delta(\omega-2\omega_k).
\end{align}

\section{Ranman spectra in the N\'{e}el antiferromagnetic phase of the anisotropic $J_1$-$J_2$-$J_3$ model}\label{Sec:ResultNeel}

%\subsection{Effects of the anisotropic exchange interaction}

\begin{figure}[t!]
\centering\includegraphics[%scale=0.28
width=80mm %, trim=5 130 5 20,clip %left, down, right, up
]{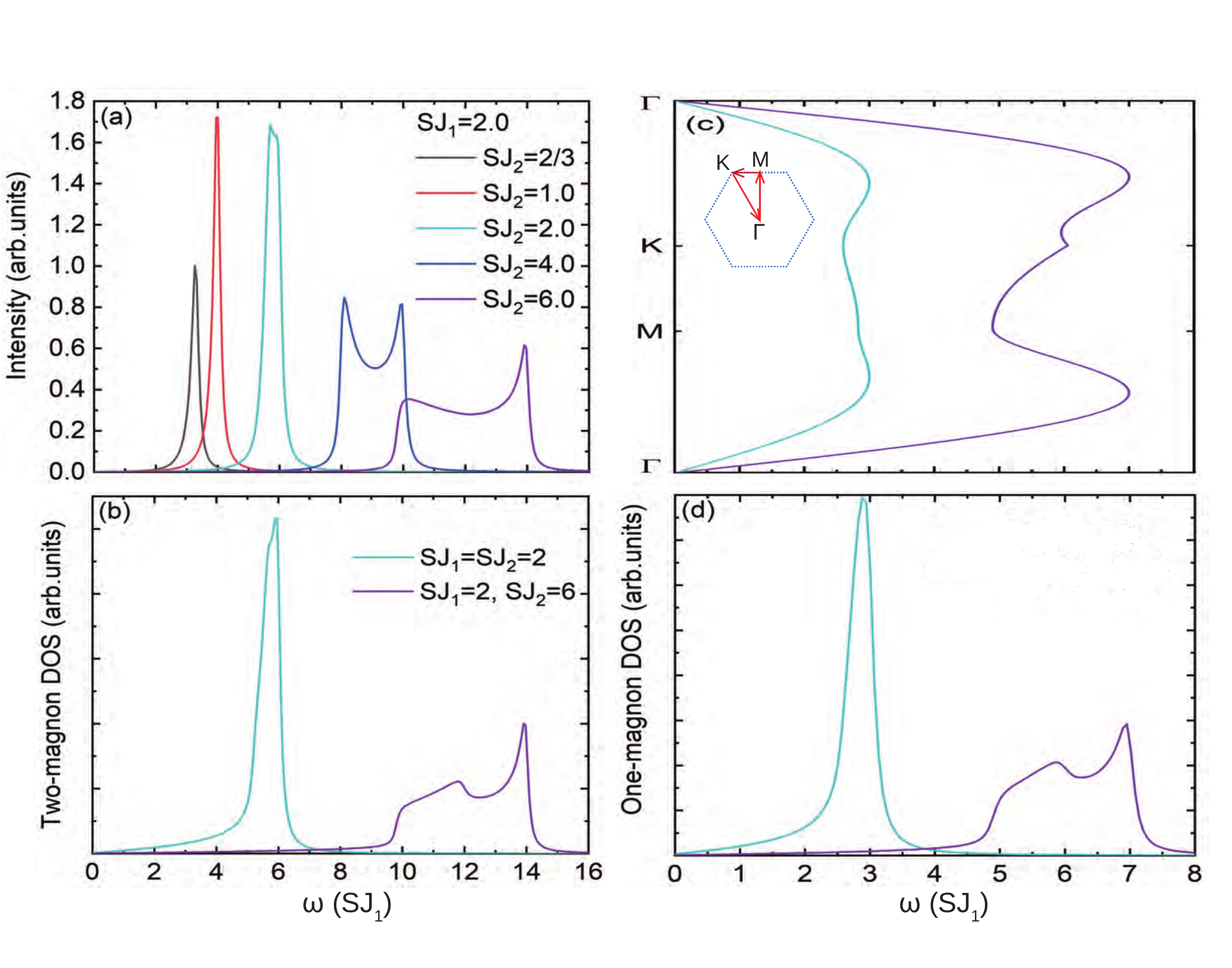}
\caption{(a): Two-magnon Raman spectra of the $J_1$-$J_2$-$J_3$ model with $J_3=0$, $SJ_1=2$, $S=2$ and several different $J_2$. (b): Corresponding two-magnon DoS. (c): Corresponding magnon dispersion. (d): Corresponding one-magnon DoS. The green and purple solid lines correspond to the results of $SJ_2=2$ and $SJ_2=6$, respectively.}
\label{fig:2}
\end{figure}

In this section we show the results in the N\'{e}el AFM state of the anisotropic $J_1$-$J_2$-$J_3$ model. We set $J_1=1$ to be the energy unit of the model, and consider the case of general spin-$S$. First we study the effects of the spatial anisotropic interaction on the Raman spectrum by setting $J_3=0$ and varying $J_2$ in the model. The calculated two-magnon Raman spectra are shown in Fig.~\ref{fig:2}(a). It is found that when $J_2\lesssim J_1$ the spectrum is narrow with a sharp peak. But for $J_2>J_1$, with increasing $J_2$ the spectrum becomes broader and shows a clear double-peak feature. From Eq.~\eqref{Eq:Rexp} we see that the Raman spectrum is closely related to the two-magnon density of states (DoS) $\sum_k \delta(\omega-2\omega_k)$. We plot the two-magnon DoS in Fig.~\ref{fig:2}(b) for several $J_2$ values and indeed the peaks in the Raman spectrum are associated with the van Hove sigularities in the two-magnon density of states. In particular, the low-energy peak of the spectrum originates from the sharp edge of the DoS.

To understand how the peaks evolve, we further show the calculated magnon dispersion along the high-symmetry directions of the first Brillouin zone (FBZ) and corresponding one-magnon DoS in Fig.~\ref{fig:2}(c) and (d). It is quite obvious that for $J_2=6$ the edge of the DoS comes from the local minimum of the dispersion at M point, while the sharp peak at $\omega\approx 7SJ_1$ is associated with the local maximum in the dispersion. Interestingly, there is another weak peak in DoS at $\omega\approx 6SJ_1$ coming from the saddle point near the K point. But it does not contribute significantly to the Raman spectrum due to the coherent factor $M_k$.

When reducing the $J_2/J_1$ ratio, it is found that the dispersion between the K and M points becomes flatter, and the dispersion at the M point turns to a saddle point. This flat part of the dispersion leads to a large DoS, which accounts for the sharp peak in the Raman spectrum. From Eq.~\eqref{Eq:PQ}, we see that this effect is strong around $J_2\approx J_1$, where the anisotropy of the interactions is weak. Therefore, the lineshape of the Raman spectrum provides valuable information of the interaction anisotropy associated with the lattice distortion in the system.

%\subsection{Effects of frustration}

\begin{figure}[t!]
\centering\includegraphics[%scale=0.28
width=80mm %, trim=5 130 5 20,clip %left, down, right, up
]{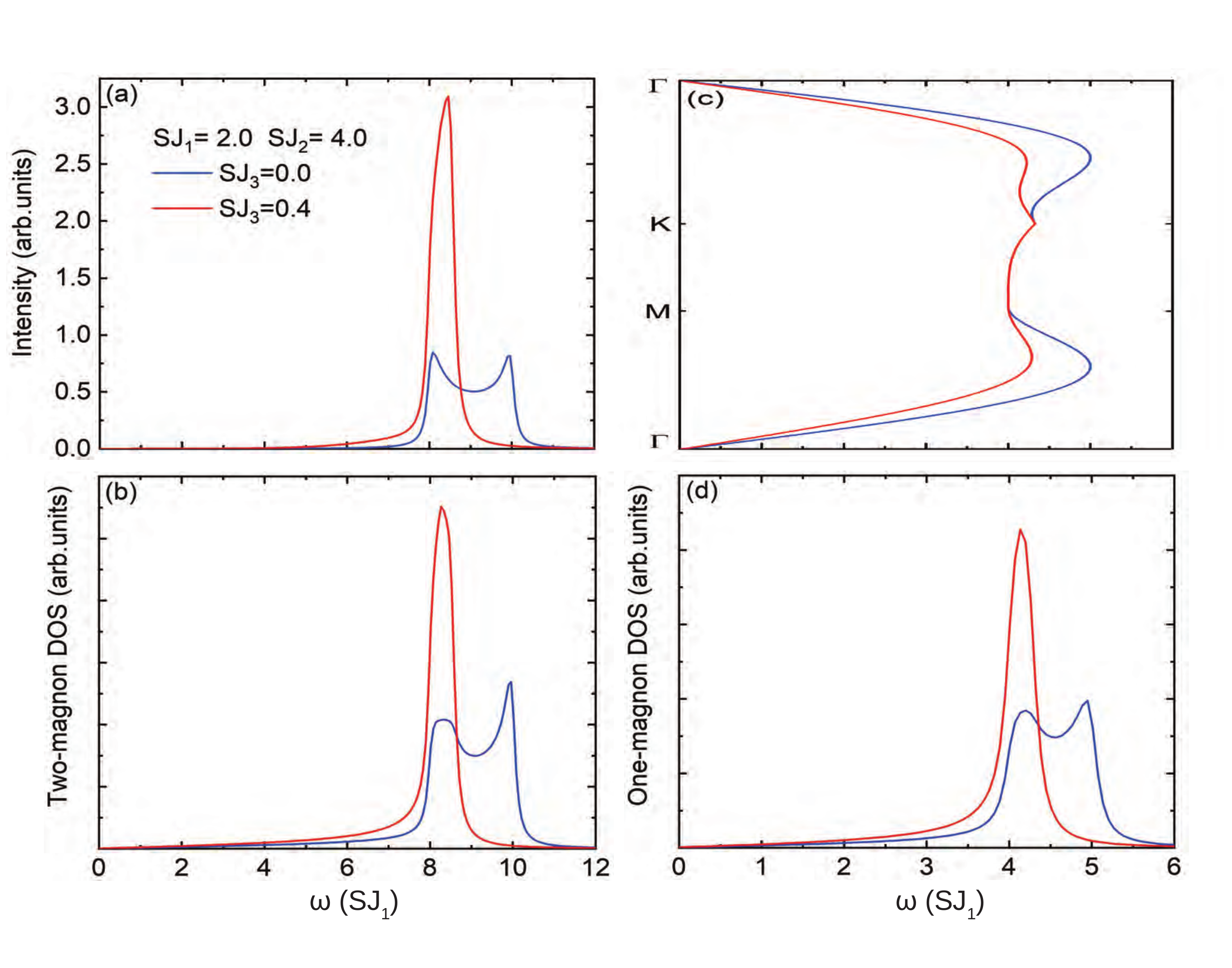}
\caption{(a): Two-magnon Raman spectra of the $J_1$-$J_2$-$J_3$ model with $SJ_1=2$, $SJ_2=4$, $S=2$ and two different $J_3$ values, showing the effects of spin frustration. (b): Corresponding two-magnon DoS. (c): The corresponding magnon dispersions. (d): Corresponding one-magnon DoS. The blue and red solid lines refer to the results of $SJ_3=0$ and $SJ_3=0.4$, respectively.}
\label{fig:3}
\end{figure}

We then study the effects of the next-nearest neighboring interaction $J_3$. In Fig.~\ref{fig:3}(a) we compare the Raman spectra with and without the frustrated $J_3$ coupling. Interestingly, the frustrated coupling can also turn the double peak structure of the spectrum to a sharp single peak. However, by comparing the dispersions in the two cases in Fig.~\ref{fig:3}(c), we see that this sharpening of the spectrum has a different origin. Note that $\gamma_{k3}=\cos (\sqrt{3}k_y)$, therefore, a finite $J_3$ does not modify the dispersion along K to M in the FBZ. However, the frustrated coupling largely suppresses the local maxima along $\Gamma$-to-K and $\Gamma$-to-M directions. This makes the dispersion almost flat-topped which contribute a sharp peak in the magnon DoS, as shown in Fig.~\ref{fig:3}(d). This large DoS gives rise to a sharp peak in the Raman spectrum, which may also turn the double peak of the spectrum to a single peak. However, different from the case of anisotropic interaction, here the energy of the peak only weakly depends on $J_3$ because it is mainly determined by the energy minimum along M-to-K direction.

\section{Ranman spectra in the N\'{e}el and stripe antiferromagnetic phases -- role of the Kitaev interaction}\label{Sec:ResultKitaev}

\begin{figure}[t!]
\centering\includegraphics[%scale=0.28
width=80mm %, trim=5 130 5 20,clip %left, down, right, up
]{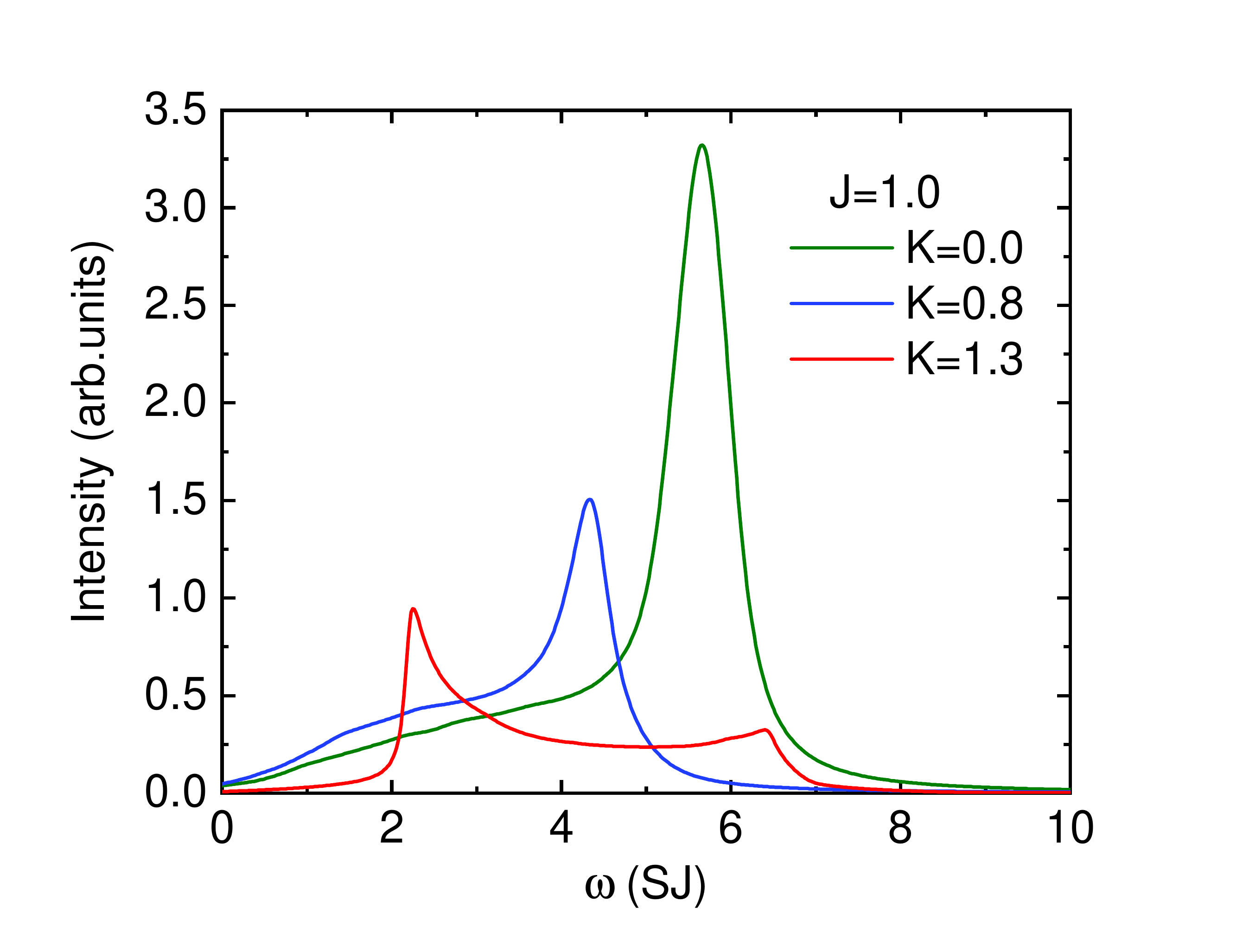}
\caption{Two-magnon Raman spectra for the $J$-$K$ model in Eq.~\eqref{Eq:M2} with $J=1$, $S=2$, and several different $K$. The green,blue and red solid lines correspond to the results of $K=0$, $K=0.8$, and $K=1.3$, respectively.}
\label{fig:4}
\end{figure}

In this section, we show that the Raman scattering spectrum is also sensitive to the spin anisotropic interactions, such as the Kitaev couplings. The calculated Raman spectra for several different Kitaev couplings of the model in Eq.~\eqref{Eq:M2} are shown in Fig.~\ref{fig:4}. In this model, we set $J=1$ as the energy unit. The classical phase diagram of this model has been studied, and it is found that the N\'{e}el AFM state is stablized at $K/J<1$. While for $K/J>1$ the ground state has a stripe AFM order (see Fig.~\ref{fig:1}(c)). With increasing $K$, the overall spectral weight reduces. At the same time, the spectral weight transfers from high energy to low energy. In the N\'{e}el AFM state, the spectrum keeps a single-peak structure. But in the stripe ordered phase, it exhibits a clear double-peak structure. Note that the $K$ term is an Ising anisotropy in the spin space, when it increases, the transverse spin fluctuations associated with the magnon excitations is suppressed. This accounts for the reduction of the spectral weight.

\section{Implication for magnetic excitations of BaFe$_2$Se$_2$O}\label{Sec:Fe}
\begin{figure}[t!]
\centering\includegraphics[%scale=0.28
width=80mm %, trim=5 130 5 20,clip %left, down, right, up
]{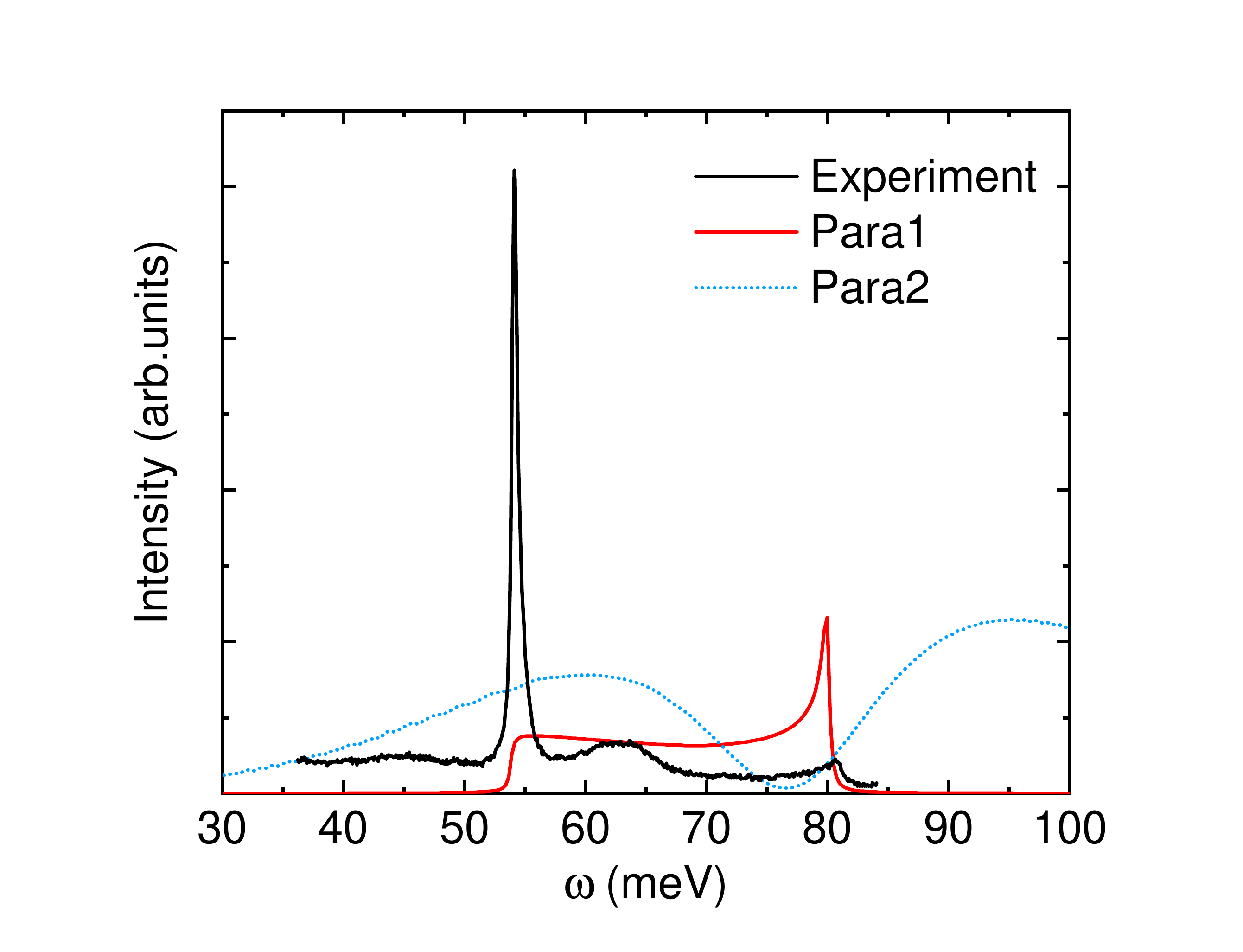}
\caption{Measured and calculated two-magnon Raman spectra for $BaFe_2Se_2O$.The black solid line corresponds to the experimental data, which was measured at 10K. We fit to the experimental data via an $S=2$ $J_1$-$J_2$-$J_3$ model with two sets of parameters: Para1 with $SJ_1=5.2$ meV, $SJ_2=17.4$ meV, $SJ_3=0$, is our best fit, and Para2 with $SJ_1=19.2$ meV, $SJ_2=31$ meV, $SJ_3=11$ meV, is taken from the first-principles calculations~\cite{Han_PRB2012}.}
\label{fig:5}
\end{figure}

In this section, we present an example on how to determine the exchange couplings by fitting to the Raman spectrum data. We consider an iron-based compound BaFe$_2$Se$_2$O, which is known as an experimental realization of coupled two-leg spin ladders~\cite{Han_PRB2012, Lei_PRB2012, Popovi_PRB2014}. In this compound, all iron ions are in the Fe$^{2+}$ oxidation state with a high spin $S$ = 2. The magnetic susceptibility shows a broad maximum at $T_{max} \approx$ 450 K and three successive magnetic phase transitions at $T \approx$ 240 K, 115 K and 43 K, respectively. $T_{max}$ signals the formation of short-range correlation among the local moments~\cite{Han_PRB2012}. The transition at $240$ K is understood as an AFM ordering, and the transitions at lower temperatures are explained as due to the formation of spin singlet dimers~\cite{Lei_PRB2012}. Lattice dynamics study of BaFe$_2$Se$_2$O
observed a magnetic excitation related structure in the form of a magnon continuum~\cite{Popovi_PRB2014}, and clear two-magnon continuum down to $10$ K has been identified in recent Raman scattering measurements~\cite{Jin_PRB2019}.

Though identified as a spin ladder compound, the effective spin model is actually defined on the honeycomb lattice, and a minimal model is just the anisotropic $J_1$-$J_2$-$J_3$ model in Eq.~\eqref{Eq:M1}. Local-spin-density-approximation (LSDA) calculation~\cite{Han_PRB2012} suggest that $J_1$ is more than three times stronger than $J_2$. With this parametrization the system would have a strong tendency to dimerization and it is unclear whether the observed magnon continuum can be understood.

Here we calculate the two-magnon Raman spectrum of the anisotropic $J_1$-$J_2$-$J_3$ model. By fitting to the experimental two-magnon spectrum data, we determine the exchange couplings. The comparison between the calculated spectra and the experimental one is shown in Fig.~\ref{fig:5}. We present results of two sets of exchange parameters. Para1 with $SJ_1=5.2$ meV, $SJ_2=17.4$ meV, $SJ_3=0$, is our best fit, while Para2 with $SJ_1=19.2$ meV, $SJ_2=31$ meV, $SJ_3=11$ meV, is taken from the LSDA calculation~\cite{Han_PRB2012}. It is clear that the spectrum with Para1 successfully reproduces the peaks at about $55$ meV and $80$ meV, while the one with Para2 does not. Note that the low-energy peak at about $55$ meV in the measured spectrum is much higher than the calculated one. This inconsistency is mainly because we do not include the effects of magnon-magnon interaction within our calculation. It is known that a sharp resonance peak slightly below the absorption edge will appear and the high-energy spectral weight will be suppressed when the magnon-magnon interaction is included~\cite{Liu_PRB2017}. Note that the determined exchange couplings from the best fit to the experimental spectrum (Para1) differ completely from those obtained in first-principles calculation (Para2). Our results suggest the system is far from dimerization and this put the previous understanding on the transitions at $115$ K and $43$ K under question. The nature of these transitions are beyond the scope of the present paper, and a detailed discussion on this issue will be present elsewhere.

\section{Conclusion}\label{Sec:Conclusion}
In conclusion, we have calculated the two-magnon Raman scattering spectra in AFM states of two frustrated spin models defined on the honeycomb lattice. We show that the spatial anisotropy, the spin anisotropy, and the frustration of the exchange interactions can all strongly affect the two-magnon spectra. More interestingly, the clear features of the two-magnon spectrum allow us to extract the completing exchange couplings via fit to the experimental data. This provides valuable information in understanding the low-energy physics of spin frustrated systems.

%\acknowledgements
\begin{acknowledgments}
We thank C. Liu, Z. X. Liu, Y. Wang, and N. Xi for useful discussions. This work has in part been supported by Ministry of Science and Technology of China,
National Program on Key Research Project Grant No.2016YFA0300504, the National Science Foundation of China Grant No. 11674392, and Research Funds of Remnin University of China Grant No. 18XNLG24 (R.Y. and J.P), and by the National Key Research and Development Program of China Grant No. 2017YFA0302904 and No. 2016YFA0300504, the National Science Foundation of China Grant No. U1932215 and No. 11774419, and the Strategic Priority Research Program of the Chinese Academy of Sciences Grant No. XDB33010100 (Q.Z., J.J., and F.J.). Q.Z. acknowledges the support from Users with Excellence Program of Hefei Science Center and High Magnetic Field Facility, CAS.
\end{acknowledgments}

%%%%%%%%%%%%%%%%%%%%%%%%%%%%%%%%%%%%%%%%%%%%%
%\bibliographystyle{abbrv}
%\bibliographystyle{unsrtnat}
%\bibliography{article}

\begin{thebibliography}{99}
\bibitem{Diep_Book} H. T. Diep, \emph{Frustrated Spin Systems}, World Scientific Publishing (2004).

\bibitem{Lacroix_Book} C. Lacroix, P. Mendels, and F. Mila, \emph{Introduction
to Frustrated Magnetism -- Materials, Experiments, Theory}, Springer-Verlag, Berlin (2011).

\bibitem{Villalba_PRB2019} M. E. Villalba, F. A. Gomez Albarracin, H. D. Rosales, D. C. Cabra, Phys. Rev. B {\bf 100}, 245106 (2019).

\bibitem{Savary_RoPP2017} L. Savary and L. Balents, ``Quantum spin liquids: a review", Rep. Prog. Phys. {\bf 80}, 016502 (2017).

\bibitem{Zhou_RMP2017} Y. Zhou, K. Kanoda, T.-K. Ng, ``Quantum spin liquid states", Rev. Mod. Phys. 89, 025003 (2017).

\bibitem{Chamorro_CR2021} J. R. Chamorro, T. M. McQueen, and T. T. Tran, ``Chemistry of Quantum Spin Liquids", Chem. Rev. {\bf 121}, 2898-2934 (2021).

\bibitem{Fu_Science2015} M. Fu {\it et al.}, ``Evidence for a gapped spin-liquid ground state in a kagome Heisenberg antiferromagnet", Science {\bf 350}, 655 (2015).

\bibitem{Shen_Nat2016} Y. Shen {\it et al.}, ``Evidence for a spinon Fermi surface in a triangular-lattice quantum-spin-liquid candidate", Nature {\bf 540}, 559 (2016).

\bibitem{Albuquerque_PRB2011} A. F. Albuquerque, D. Schwandt, B. Het\'{e}nyi, S. Capponi, M. Mambrini, and A. M. Lauchli, ``Phase diagram of a frustrated quantum antiferromagnet on
  the honeycomb lattice: Magnetic order versus valence-bond crystal
  formation", Phys. Rev. B {\bf 84}, 024406 (2011).

\bibitem{Ganesh_PRL2013} R. Ganesh, J. van den Brink, and S. Nishimoto, Phys. Rev. Lett. {\bf 110}, 127203 (2013).

\bibitem{Pujari_PRB2015} S. Pujari, F. Alet, and K. Damle, ``Transitions to valence-bond solid order in a honeycomb lattice antiferromagnet", Phys. Rev. B {\bf 91}, 104411 (2015).

\bibitem{Kitaev_AP2006} A. Kitaev, ``Anyons in an exactly solved model and beyond", Annal. Phys. {\bf 321}, 2-111 (2006).

\bibitem{Rastelli_Physica1979} E. Rastelli, A. Tassi, and L. Reatto, Physica B+C {\bf 97}, 1 (1979).

\bibitem{Fouet_EPJB2001} J. B. Fouet, P. Sindzinger, and C. Lhuillier, ``An investigation of the quantum $J_1$-$J_2$-$J_3$ model on the honeycomb lattice", Eur. Phys. J. B {\bf 20}, 241 (2001).

\bibitem{Rogado_PRB2002} N. Rogado, Q. Huang, J. W. Lynn, A. P. Ramirez, D. Huse, and R. J. Cava, ``BaNi$_2$V$_2$O$_8$: A two-dimensional honeycomb antiferromagnet", Phys. Rev. B {\bf 65}, 144443 (2002).

\bibitem{Regnault_Physica1977} L. Regnault, P.Burlet, and J. Rossat-Mignod, ``Magnetic ordering in a planar $x$-$y$ model: BaCo$_2$(AsO$_4$)$_2$", Physica B+C {\bf 86-88} 660-662 (1977).

\bibitem{Sears_PRB2015} J. A. Sears, M. Songvilay, K. W. Plumb, J. P. Clancy, Y. Qiu, Y. Zhao,
  D. Parshall, and Y.-J. Kim, ``Magnetic order in $\alpha$-RuCl$_3$: A honeycomb-lattice
  quantum magnet with strong spin-orbit coupling", Phys. Rev. B {\bf 91}, 144420 (2015).

\bibitem{Asai_PRB2017} S. Asai, M. Soda, K. Kasatani, T. Ono, V. O. Garlea, B. Winn, and T. Masuda, ``Spin dynamics in the stripe-ordered buckled honeycomb lattice antiferromagnet Ba$_2$NiTeO$_6$", Phys. Rev. B {\bf 96}, 104414 (2017).

\bibitem{Luo_SP2020} Z.-X. Luo and G. Chen, ``Honeycomb rare-earth magnets with anisotropic exchange
  interactions", Sci. Post Phys. Core {\bf 3}, 4 (2020).

\bibitem{Ji_CPL2021} J. Ji, M. Sun, Y. Cai, Y. Wang, Y. Sun, W. Ren, Z. Zhang, F. Jin, and Q. Zhang, ``Rare-earth chalcohalides: A family of van der Waals layered Kitaev spin liquid candidates", Chin. Phys. Lett. {\bf 38}, 047502 (2021).

\bibitem{Kim_PRB2016} H.-S. Kim and H.-Y. Kee, ``Crystal structure and magnetism in $\alpha$-RuCl$_3$: an ab-initio study", Phys. Rev. B {\bf 93}, 155143 (2016).

\bibitem{Sun_MRE2017} Y. C. Sun, Z. W. Ouyang, Y. Xiao, Y. Su, E. Feng, Z. Fu, W. T. Jin, M. Zbiri, Z. C. Xia, J. F. Wang, and G. H. Rao, ``Honeycomb-lattice antiferromagnet Mn$_2$V$_2$O$_7$: a temperature-dependent x-ray diffraction, neutron diffraction and ESR study", Mater. Res. Express {\bf 4}, 046101 (2017).

\bibitem{Takagi_NRP2019} H. Takagi, T. Takayama, G. Jackeli, G. Khaliullin, and S. E. Nagler, ``Concept and realization of Kitaev quantum spin liquids", Nat. Rev. Phys. {\bf 1}, 264 (2019).

\bibitem{Devereaux_RMP2007}
T. P. Devereaux and R. Hackl, ``Inelastic light scattering from correlated
  electrons", Rev. Mod. Phys. {\bf 79}, 175-233 (2007).

\bibitem{Lemmens_PR2003}
P. Lemmens, G. G{\"u}ntherodt, and C. Gros, ``Magnetic light scattering in
  low-dimensional quantum spin systems", Phys. Rep. {\bf 375}, 1-103 (2003).

\bibitem{Fleury_PRL1967}
P. Fleury, S. Porto, and R. Loudon, ``Two-magnon light scattering in
  antiferromagnetic MnF$_2$", Phys. Rev. Lett. {\bf 18} 658 (1967).

\bibitem{Fleury_PR1968} P. Fleury and R. Loudon, ``Scattering of light by one-and two-magnon
  excitations", Physical Review {\bf 166}, 514 (1968).

\bibitem{Shastry_PRL1990} B. S. Shastry and B. I. Shraiman, Phys. Rev. Lett. {\bf 65}, 1068 (1990).

\bibitem{Han_PRB2012} F. Han, X. Wan, B. Shen, and H.-H. Wen, ``BaFe$_2$Se$_2$O as an
  iron-based mott insulator with antiferromagnetic order", Phys. Rev. B {\bf 86}, 014411 (2012).

\bibitem{Popovi_PRB2014} Z. V. Popovi {\it et al.}, ``Phonon and magnetic dimer excitations in Fe-based $S=2$ spin-ladder compound BaFe$_2$Se$_2$O", Phys. Rev. B {\bf 89}, 014301 (2014).

\bibitem{Jin_PRB2019} F. Jin {\it et al.}, ``Phonon anomalies and magnetic excitations in
  BaFe$_2$Se$_2$O", Phys. Rev. B {\bf 99}, 144419 (2019).

\bibitem{Price_PRB2013} C. Price and N. B. Perkins, ``Finite temperature phase diagram of the classical Kitaev-Heisenberg model", Phys. Rev. B {\bf 88}, 024410 (2013).

\bibitem{Stavropoulos_PRL2019} P. P. Stavropoulos, D. Pereira, and H.-Y. Kee, ``Microscopic mechanism for higher-spin Kitaev model", Phys. Rev. Lett. {\bf 123}, 037203 (2019).

\bibitem{Consoli_PRB2020} P. M. C\^{o}nsoli, L. Janssen, M. Vojta, and E. C. Andrade, ``Heisenberg-Kitaev model in a magnetic field: $1/S$ expansion", Phys. Rev. B {\bf 102}, 155134 (2020).

\bibitem{Lei_PRB2012} H. Lei, H. Ryu, V. Ivanovski, J. B. Warren, A. I. Frenkel, B. Cekic, W.-G. Yin, and C. Petrovic, ``Structure and physical properties of the layered iron oxychalcogenide BaFe$_2$Se$_2$O", Phys. Rev. B {\bf 86}, 195133 (2012).

\bibitem{Liu_PRB2017} C. Liu, A. Zhang, Q. Zhang, R. Yu, and X. Wang, ``Spin-wave approach to the
  two-magnon raman scattering in a $J_{1x}$-$J_{1y}$-$J_2$-$J_c$ antiferromagnetic
  heisenberg model", Physical Review B {\bf 95}, 104431 (2017).
%%%%%%%%%%%%%%%%%%%%%%%%%%%%%%%%%%%



\end{thebibliography}

\end{document}